\documentclass[twocolumn,showpacs,prl,aps,superscriptaddress]{revtex4}

\usepackage{amsmath}
\usepackage{amssymb}
\usepackage{graphicx}
\usepackage{upgreek}
\usepackage{bm}

\newcommand{\vect}[1]{\bm{#1}}

\newcommand{\dif}{\mathrm{d}}
\newcommand{\mi}{\mathrm{i}}
\newcommand{\me}{\mathrm{e}}

\newcommand{\jsum}{\sum_{j=0}^\infty \!{}^{{}^\prime}}

\newcommand{\zA}{z_\mathrm{A}}

\newcommand{\be}{\begin{equation}}
\newcommand{\ee}{\end{equation}}

\newcommand{\re}{\mathrm{Re}}

\begin{document}

\title{Temperature-invariant Casimir--Polder forces despite
large thermal photon numbers}

\author{Simen {\AA}. Ellingsen}
\affiliation{Department of Energy and Process Engineering, Norwegian
University of Science and Technology, N-7491 Trondheim, Norway}
\email{simen.a.ellingsen@ntnu.no}
\author{Stefan Yoshi Buhmann}
\author{Stefan Scheel}
\affiliation{Quantum Optics and Laser Science, Blackett Laboratory,
Imperial College London, Prince Consort Road,
London SW7 2AZ, United Kingdom}

\date{\today}

\begin{abstract}
We demonstrate that Casimir--Polder potentials can be entirely
independent of temperature even when allowing for the thermal photon
numbers to become large at the relevant molecular transition
frequencies. This statement holds for potentials that are due to
low-energy transitions of a molecule placed near a plane metal
surface. For a molecule in an energy eigenstate, the
temperature-invariance is a consequence of strong cancellations
between nonresonant potential components and those due to evanescent
waves. For a molecule with a single dominant transition in a thermal
state, upward and downward transitions combine to a
temperature-independent potential. The results are contrasted with the
case of an atom whose potential exhibits a regime of linear
temperature-dependence. Contact to the Casimir force between a weakly
dielectric and a metal plate is made.
\end{abstract}

\maketitle

Dispersion forces between polarisable objects were originally
predicted by Casimir and Polder as a consequence of quantum zero-point
fluctuations \cite{CasimirPolder,Casimir}. Recent measurements of both
Casimir--Polder (CP) forces between atoms and surfaces \cite{Sukenik}
and Casimir forces between macroscopic bodies \cite{capasso07,bordag}
typically operate at room temperature where thermal fluctuations also
come into play \cite{lifshitz55,nakajima97,gorza06,buhmann08}. The
temperature-dependence of dispersion forces is of relevance for both
fundamental and practical reasons.

On the theoretical side, the correct description of the Casimir force
between metals at finite temperature is subject to an ongoing debate
\cite{bordag,finiteCasimir}. To wit, predictions differ for the
high-temperature behaviour of the Casimir force between metals, for
which employing a standard dissipative description of the surfaces 
fails to reproduce the experimental observations \cite{decca}. This
suggests that progress can be made by directly observing the variation
of the Casimir force with temperature.

On the practical side, CP forces become increasingly relevant when
trying to trap and coherently manipulate cold atoms near surfaces
\cite{Vuletic}. Current endeavours aim at extending these techniques
to more complex systems such as polar molecules \cite{vanfeldhoven}.
Such systems typically exhibit long-wavelength transitions so that
CP forces become increasingly long-ranged. This raises the question
whether they can be controlled by lowering the ambient
temperature and hence suppressing thermal force components.

Thermal contributions to the CP potential are governed by the 
photon number $n(\omega)=[\me^{\hbar\omega/(k_\mathrm{B}T)}-1]^{-1}$.
A noticeable deviation of the potential from its zero-temperature
value is to be expected when $n(\omega)\!\gtrsim\!1$ in the relevant
frequency range. This is the case, for instance, for molecules with
small transition frequencies,
$|\omega_{kn}|\!\lesssim\!k_\mathrm{B}T/\hbar\!=\!3.93\times
10^{13}\,\mathrm{rad/s}$ at room temperature ($300\,\mathrm{K}$).
The associated wavelengths are much larger than typical experimental
molecule--surface separations in the nanometre to micrometre range 
\cite{Sukenik}, $\zA\!\ll\!c/|\omega_{kn}|$. Furthermore, experimental
realisations typically involve conducting and thus highly reflecting
metal surfaces with $|\varepsilon(\omega_{kn})|\!\gg\!1$.

As we will show in this Letter, the above three conditions
combined result in potentials which are independent of temperature
over the entire range from zero to room temperature and beyond. We
will first discuss the case of a molecule prepared in an energy
eigenstate and then consider molecules at thermal equilibrium with
their environment, comparing our results with those for atoms whose 
transitions involve higher energies.


\paragraph{Molecule vs.\ atom in an eigenstate.}

As shown in Ref.~\cite{ellingsen09a}, the CP potential of a molecule
prepared in an isotropic energy eigenstate $|n\rangle$ at distance
$\zA$ from the plane surface of a metal,
\begin{equation}
\label{ti2}
U_n(\zA)=U_n^\mathrm{nr}(\zA)
 +U_n^\mathrm{ev}(\zA)+U_n^\mathrm{pr}(\zA),
\end{equation}
naturally separates into three contributions: a non-resonant term
$U_n^\mathrm{nr}$ due to virtual photons that is formally similar to
that produced by Lifshitz theory \cite{lifshitz55}, and a resonant
contribution due to real photons which may be further split into
contributions from evanescent ($U_n^\mathrm{ev}$) and propagating
($U_n^\mathrm{pr}$) waves. The non-resonant potential is given by
\cite{ellingsen09a}
\begin{align}
  U&_n^\mathrm{nr}(\zA)
  =-\frac{\mu_0k_\mathrm{B}T}{6\pi\hbar}\sum_k|\vect{d}_{nk}|^2
  \jsum \frac{\omega_{kn}}{\omega_{kn}^2
 +\xi_j^2}\int_{\xi_j/c}^\infty\dif b \notag\\
&\times\me^{-2b\zA}
 \bigl\{2b^2c^2r_p(\mi\xi_j)-\xi_j^2[r_s(\mi\xi_j)+r_p(\mi\xi_j)]\bigr\} 
  \label{ti3}
\end{align}
[$\omega_{kn}\!=\!(E_k\!-\!E_n)/\hbar$, transition frequencies;
$\vect{d}_{nk}$, dipole matrix elements;
$\xi_j=j\xi$ with $\xi=2\pi k_\mathrm{B}T/\hbar$, Matsubara
frequencies; the primed summation indicates that the term $j=0$
carries half weight] and the evanescent one reads
\begin{multline}
\label{ti4}
U_n^\mathrm{ev}(\zA)
=\frac{\mu_0}{12\pi}\sum_kn(\omega_{kn})|\vect{d}_{nk}|^2
 \int_0^\infty\dif b\,\me^{-2b\zA}\bigl\{2b^2c^2\\
\times\re[r_p(\omega_{kn})]
 +\omega_{nk}^2\re[r_s(\omega_{kn})+r_p(\omega_{kn})]\bigr\},
\end{multline}
note that $n(\omega_{kn})\!=\!-[n(\omega_{nk})+1]$ for downward
transitions. These two contributions dominate in the region
$\zA|\omega_{kn}|/c\!\ll\!1$ we are interested in, while the spatially
oscillating $U_n^\mathrm{pr}$ becomes relevant only in the far-field
range $\zA|\omega_{kn}|/c\!\gg\!1$. The reflection coefficients of
the surface for $s$- and $p$-polarised waves are given by
$r_s(\omega)\!=\!(b-b_1)/(b+b_1)$
and $r_p(\omega)\!=\!%
[\varepsilon(\omega)b-b_1]/[\varepsilon(\omega)b+b_1]$ with
$b_1\!=\!\sqrt{b^2-[\varepsilon(\omega)-1]\omega^2/c^2}$,
$\re(b_1)\!>\!0$.

For a metal surface whose plasma frequency is typically much larger
than the molecular transition frequency $\omega_{kn}$, we have
$|\varepsilon(\omega)|\!\gg\!1$ in the relevant frequency range, so
that the reflection coefficients $r_s\!\approx\!-1$ and
$r_p\!\approx\!1$ become frequency-independent. The $b$-integrals can
then be performed to give
\begin{align}
\label{ti6}
U_n^\mathrm{nr}(\zA)
=&-\frac{k_\mathrm{B}T}{12\pi\varepsilon_0\hbar \zA^3}
 \sum_k|\vect{d}_{nk}|^2
 \jsum \frac{\omega_{kn}\me^{-2j\zA\xi/c}}
 {\omega_{kn}^2+j^2\xi^2}\nonumber\\
&\times
 \biggl[1+2j\,\frac{\zA\xi}{c}+2j^2\,\frac{\zA^2\xi^2}{c^2}
 \biggr],\\
\label{ti7}
U_n^\mathrm{ev}(\zA)
=&\frac{1}{24\pi\varepsilon_0\zA^3}
 \sum_kn(\omega_{kn})|\vect{d}_{nk}|^2.
\end{align}

The asymptotic temperature-dependence of the potential for
a given distance from the surface is governed by two characteristic
temperatures: The molecular transition frequency defines a
spectroscopic temperature
$T_\omega\!=\!\hbar|\omega_{kn}|/k_\mathrm{B}$, which is roughly the
temperature required to noticeably populate the upper level.
Similarly, the distance introduces a geometric temperature
$T_z\!=\!\hbar c/(\zA k_\mathrm{B})$, i.e., the temperature of
radiation whose wavelength is of the order $\zA$. 

\begin{figure}[t]
\includegraphics[width=2.9in]{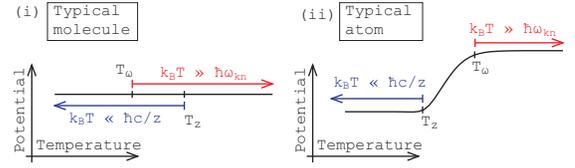}
\caption{\label{Fig1}
Sketch of the temperature-dependence of the CP potential for a typical
molecule vs.\ a typical atom.}
\end{figure}

We will now show that the total potential becomes independent of
temperature in both the geometric low-temperature limit $T\!\ll\!T_z$
and the spectroscopic high-temperature limit $T\!\gg\!T_\omega$. For a
typical molecule with its long-wavelength transitions, the potential
is nonretarded for typical molecule--surface distances,
$\zA|\omega_{kn}|/c\!\ll\!1$. As depicted in Fig.~\ref{Fig1}(i), this
implies $T_\omega\!\ll\!T_z$, hence the two regions of constant
potential overlap and the potential is constant for all temperatures.
For an atom, on the contrary, the transition wavelengths are much
shorter, so that we may have $\zA|\omega_{kn}|/c\!\gg\!1$. In this
case, an intermediate regime $T_z\!\ll\!T\!\ll\!T_\omega$ exists where
the potential increases linearly with temperature,
cf.~Fig.~\ref{Fig1}(ii).

We begin with a typical molecule with $\zA|\omega_{kn}|/c\!\ll\!1$. In
the geometric low-temperature limit $T\!\ll\!T_z$, we have
$\zA\xi/c\!\ll\!1$, hence the sum in Eq.~(\ref{ti6}) is densely
spaced. The factor $1/(\omega_{kn}^2+j^2\xi^2)$ restricts it to values
where $j\zA\xi/c\!\le\!\zA|\omega_{kn}|/c\!\ll\!1$. With this
approximation, the summation can be performed as
\begin{equation}
\label{ti8}
\jsum\frac{1}{a^2+j^2}
 =\frac{\pi}{2a}\,\coth(\pi a)
\end{equation}
and we find
\begin{equation}
\label{ti9}
U_n^\mathrm{nr}(\zA)
\!\!\buildrel{\scriptscriptstyle T\ll T_z}\over{=}\!\!
-\frac{1}{24\pi\varepsilon_0 \zA^3}
 \sum_k\bigl[n(\omega_{kn})+{\textstyle\frac{1}{2}}\bigr]
 |\vect{d}_{nk}|^2,
\end{equation}
noting that
$\coth[\hbar\omega_{kn}/(2k_\mathrm{B}T)]\!=\!2n(\omega_{kn})\!+\!1$.
Adding the evanescent contribution~(\ref{ti7}), we find the
temperature-independent total potential
\begin{equation}
\label{ti10}
U_n(\zA)
\!\!\buildrel{\scriptscriptstyle T\ll T_z}\over{=}\!\!
-\frac{\sum_k|\vect{d}_{nk}|^2}
{48\pi\varepsilon_0\zA^3}
=-\frac{\langle\hat{\vect{d}^2}\rangle_n}
 {48\pi\varepsilon_0\zA^3}\,,
\end{equation}
in agreement with the well-known nonretarded zero-temperature result
\cite{CasimirPolder}.

In the spectroscopic high-temperature limit $T\!\gg\!T_\omega$, we
have $\xi/|\omega_{kn}|\!\gg\!1$. Due to the denominator
$\omega_{kn}^2+j^2\xi^2$, the $j=0$ term strongly dominates the sum
in Eq.~(\ref{ti6}) and we find
\begin{equation}
\label{ti11}
U_n^\mathrm{nr}(\zA)
\!\!\buildrel{\scriptscriptstyle T\gg T_\omega}\over{=}\!\!
-\frac{1}{24\pi\varepsilon_0 \zA^3}
 \sum_k\frac{k_\mathrm{B}T}{\hbar\omega_{kn}}\,|\vect{d}_{nk}|^2.
\end{equation}
Under the condition $T\!\gg\! T_\omega$, i.e.,
$k_\mathrm{B}T\!\gg\!\hbar|\omega_{kn}|$, the evanescent
contribution~(\ref{ti7}) reduces to
\begin{equation}
\label{ti12}
U_n^\mathrm{ev}(\zA)
\!\!\buildrel{\scriptscriptstyle T\gg T_\omega}\over{=}\!\!
\frac{1}{24\pi\varepsilon_0\zA^3}
 \sum_k\biggl(\frac{k_\mathrm{B}T}{\hbar\omega_{kn}}
 -\frac{1}{2}\biggr)|\vect{d}_{nk}|^2.
\end{equation}
Adding the two results, the total potential is again
temperature-independent and given by Eq.~(\ref{ti10}). This limit is
just the high-temperature saturation already pointed out in
Refs.~\cite{ellingsen09a,gorza06}.

\begin{figure}[t]
\includegraphics[width=2.7in]{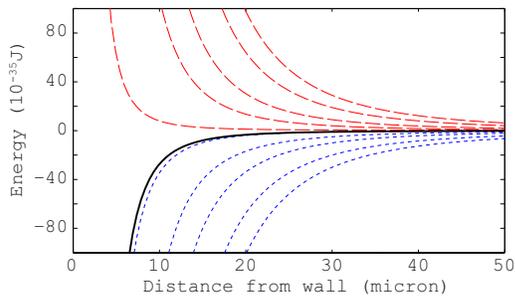}
\caption{
\label{Fig2}
CP potential of a ground-state LiH molecule in front of a Au surface.
We show the total potential (solid line) as well as its evanescent
(dashed) and nonresonant (dotted) contributions for temperatures
$10\mathrm{K}$, $50\mathrm{K}$, $100\mathrm{K}$, $200\mathrm{K}$,
$300\mathrm{K}$ (left to right).}
\end{figure}

The thermal CP potential of a typical molecule with its
long-wavelength transitions has thus been found to be
temperature-invariant in the geometric low-temperature and
spectroscopic high-temperature regimes. Due to the condition
$\zA|\omega_{kn}|/c\!\ll\!1$, at least one of the conditions  
$T\!\ll\!T_z$ or $T\!\gg\!T_\omega$ always holds, implying that the
potential is constant for all temperatures and it agrees with its
zero-temperature value. The invariance of the total potential in both
regimes is a result of cancellations between nonresonant and
evanescent potential components, which both strongly depend on
temperature. This is illustrated in Fig.~\ref{Fig2} where we display
the total temperature-invariant potential as well as its nonresonant
and evanescent parts for a ground-state LiH molecule in front of a Au
surface for various temperatures. It is seen that very strong
cancellations occur, especially at high temperatures.

We now turn to the case of atoms, whose electronic wavelengths are
short compared to typical experimental separations:
$\zA|\omega_{kn}|/c\!\gg\!1$. The exponential restricts the sum in
Eq.~(\ref{ti6}) to terms with
$j\xi\!\lesssim\!c/\zA\!\ll\!|\omega_{kn}|$, 
so the term $j^2\xi^2$ in the denominator may be neglected. The sum
can then be performed according to
\begin{multline}
\label{ti13}
\jsum\me^{-2ja}(1+2ja+2j^2a^2)=\frac{\coth(a)}{2}\\
 +\frac{a}{2}\,\frac{1+a\coth(a)}{\sinh^2(a)}
 \to\left\{{\begin{array}{cl}3/(2a),&a\ll 1;\\ 
   1/2,&a\gg 1.\end{array}}\right.
\end{multline}
In the geometric low-temperature regime, $T\!\ll\!T_z$, we further
have $\zA\xi/c\!\ll\!1$, i.e., $a\!\ll\!1$ in Eq.~(\ref{ti13}),
hence
\begin{equation}
\label{ti14}
U_n^\mathrm{nr}(\zA)
\!\!\buildrel{\scriptscriptstyle T\ll T_z}\over{=}\!\!
-\frac{c}{16\pi^2\varepsilon_0\zA^4}
 \sum_k\frac{|\vect{d}_{nk}|^2}{\omega_{kn}}\,,
\end{equation}
in agreement with the famous zero-temperature result of Casimir and
Polder \cite{CasimirPolder}. Moreover, the condition
$\zA|\omega_{kn}|/c\!\gg\!1$ implies that
$T\!\ll\!T_z\!\ll\!T_\omega$: The geometric low-temperature regime is
also a spectroscopic one and hence the evanescent potential reduces to
\begin{equation}
\label{ti15}
U_n^\mathrm{ev}(\zA)
\!\!\buildrel{\scriptscriptstyle T\ll T_z}\over{=}\!\!
-\frac{1}{24\pi\varepsilon_0\zA^3}
 \sum_k\Theta(\omega_{nk})|\vect{d}_{nk}|^2.
\end{equation}
[$\Theta(x)$, unit step function]. Adding the two contributions, we
find the temperature-independent total potential
\begin{equation}
\label{ti16}
U_n(\zA)
\!\!\buildrel{\scriptscriptstyle T\ll T_z}\over{=}\!\!
-\frac{1}{24\pi\varepsilon_0\zA^3}
\sum_k\biggl[\frac{3c}{2\zA\omega_{kn}}
+\Theta(\omega_{nk})\biggr]|\vect{d}_{nk}|^2
\end{equation}
in the geometric low-temperature regime.

For intermediate temperatures $T_z\!\ll\!T\!\ll\!T_\omega$ we have
$\zA\xi/c\!\ll\!1$, so using $a\!\gg\!1$ in the sum~(\ref{ti13}), the
nonresonant potential~(\ref{ti6}) is found to read as in
Eq.~(\ref{ti11}). The evanescent contribution is still given by
Eq.~(\ref{ti15}), so the total potential varies linearly with
temperature,
\begin{equation}
\label{ti17}
U_n(\zA)
\!\!\buildrel{\scriptscriptstyle
T_z\!\ll\!T\!\ll\!T_\omega}\over{=}\!\!
-\frac{1}{24\pi\varepsilon_0\zA^3}
 \sum_k\biggl[\frac{k_\mathrm{B}T}{\hbar\omega_{kn}}
 +\Theta(\omega_{nk})\biggr]|\vect{d}_{nk}|^2.
\end{equation}

In the spectroscopic high-temperature limit $T\!\gg\!T_\omega$, the
evanescent contribution is given by Eq.~(\ref{ti12}) as already shown. It
cancels with the nonresonant contribution, still agreeing with
Eq.~(\ref{ti11}), to give a saturated potential of the
form~(\ref{ti10}). However, with, e.g., $T_\omega\approx 18.000$~K for
Rb, this saturation is unobservable. Moreover, as the electronic
transition frequencies can be comparable to the plasma frequency of
the metal, the assumptions $r_s\!\approx\!-1$ and $r_p\!\approx\!1$ do
not hold. As a consequence, the cancellations required to achieve
saturation do not occur for atoms near realistic metal surfaces.

\begin{figure}[t]
\includegraphics[width=2.7in]{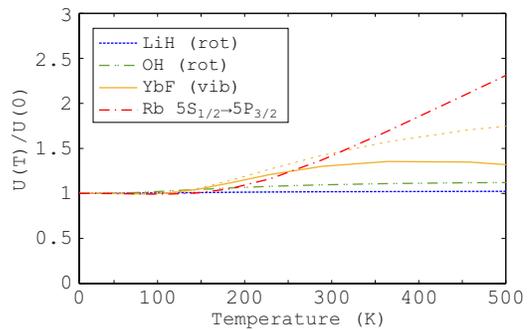}
\caption{
\label{Fig3}
Temperature-dependence of the CP potential of various ground-state
atoms and molecules at distance $\zA=5\mu\mathrm{m}$ from a Au
surface. The transition frequencies of these species are such that
$\zA\omega_{kn}/c\!=\!0.046$ (LiH), $0.26$ (OH), $1.59$ (YbF), 
$40.2$ (Rb). For comparison, the perfect-conductor result for YbF is
also shown (dotted line).}
\end{figure}

We have thus seen that for an atom with $\zA|\omega_{kn}|/c\!\gg\!1$,
separate geometric low-temperature and spectroscopic high-temperature
regimes exist, with the potential exhibiting a linear
temperature-dependence between these two regions. The difference
between the thermal CP potentials of typical atoms vs.\
molecules is illustrated in Fig.~\ref{Fig3} where we show the
temperature-dependence of the potential at fixed distance from a Au
surface for different species.
%
The potentials associated with the long-wavelength, rotational
transitions of LiH and OH are virtually temperature-invariant while
the short-wavelength electronic transition of Rb shows a linear
increase over a large range of temperatures. YbF with its dominant
vibrational transition lies in between the two extremes of typical
long-wavelength molecular and short-wavelength atomic transitions;
its potential increases by about $30\%$ in the displayed temperature
range. In contrast to the other examples, the potential of YbF
noticeably deviates from the corresponding ideal conductor-result
due to imperfect reflection. Note that contributions to the molecular
CP potentials due to electronic transitions are smaller than the
rotational and vibrational ones (\ref{ti10}) by factors
$c/(\zA\omega_{kn})\!\ll\!1$ (\ref{ti16}) or
$k_\mathrm{B}T/(\hbar\omega_{kn})\ll 1$ (\ref{ti17}) within the
displayed temperature range and are hence negligible.


\paragraph{Molecule at thermal equilibrium.} 

The proven temper\-ature-invariance immediately generalises to
molecules in incoherent superpositions of energy eigenstates with
temperature-independent probabilities $p_n$ and total potential
$U(\zA)\!=\!\sum_np_nU_n(\zA)$. The case of a molecule at thermal
equilibrium with its environment needs to be treated separately since
the respective probabilities $p_n\!=\!\exp[-E_n/(k_\mathrm{B}T)]/%
\sum_k\exp[-E_k/(k_\mathrm{B}T)]$ depend on $T$. At thermal
equilibrium, all resonant potential components cancel pairwise
\cite{buhmann08}. Introducing potential components $U_{nk}$ due to a
particular transition $n\!\leftrightarrow\!k$ (such that
$U_n\!=\!\sum_kU_{nk}$) and the associated statistical weights
$p_{nk}\!=\!p_n+p_k$, and exploiting the fact that
$U_{kn}\!=\!-U_{nk}$, we can write the total potential in the form
\begin{align}
U(\zA)=
&\sum_{n<k}(p_n-p_k)U_{nk}^\mathrm{nr}(\zA)\notag\\
=&
\sum_{n<k}p_{nk}
 \tanh\biggl(\frac{\hbar\omega_{kn}}{2k_\mathrm{B}T}\biggr)
 U_{nk}^\mathrm{nr}(\zA).
\label{ti18}
\end{align}

The behaviour of this potential in the two limits relevant for a
molecule with $\zA|\omega_{kn}|/c\!\ll\!1$ follow immediately from the
asymptotes given in the previous section. For $T\!\ll\!T_z$,
$U_{nk}^\mathrm{nr}$ from Eq.~(\ref{ti9}) leads to
\begin{equation}
\label{ti19}
U(\zA)
\!\!\buildrel{\scriptscriptstyle T\ll T_z}\over{=}\!\!
-\frac{1}{48\pi\varepsilon_0 \zA^3}
 \sum_{n<k}p_{nk}|\vect{d}_{nk}|^2,
\end{equation}
where $n(\omega_{kn})\!+\!1/2\!=\!%
\coth[\hbar\omega_{kn}/(2k_\mathrm{B}T)]/2$ has been used once more.
For $T\!\gg\!T_\omega$, we recall $U_{nk}^\mathrm{nr}$ from
Eq.~(\ref{ti11}) and note that 
$\tanh[\hbar\omega_{kn}/(2k_\mathrm{B}T)]\!\approx\!%
\hbar\omega_{kn}/(2k_\mathrm{B}T)$ to again find the
potential~(\ref{ti19}).

Combining the two results, we may use the main argument of the
previous section to conclude that the potential components associated
with a particular transition $n\!\leftrightarrow\!k$ are independent
of temperature for all temperatures. The invariance is a result of
cancellations between the purely nonresonant contributions from lower
state $n$ and upper state $k$. For larger thermal photon numbers,
these cancellations become stronger and hence counteract the increase
of the potential one might have expected. Note however that the
statistical weights $p_{nk}$ introduce a weak temperature-dependence
in general: The total potential is only strictly temperature-invariant
when dominated by a single transition.


\paragraph{Relevance to Casimir forces.}

To illustrate the relevance of the demonstrated temperature-invariance
to the Casimir force, let us consider an infinite dielectric half
space filled with molecules of number density $\eta$ at a distance $z$
from a metal plane. For a weakly dielectric medium, the Casimir energy
per unit area is given by 
$E(z)=\int_{z}^\infty\dif \zA\eta U(\zA)$ \cite{raabe06}. Using
Eq.~(\ref{ti19}), we find that 
\begin{equation}
\label{ti20}
E(z)
=\frac{\eta}{96\pi\varepsilon_0 z^2}
 \sum_{n<k}p_{nk}|\vect{d}_{nk}|^2
\end{equation}
which is temperature-independent under the conditions mentioned above.

For dielectrics with a stronger response, manybody-effects will lead
to temperature-dependent corrections of higher order in the molecular
polarisability. Such corrections are suppressed in the spectroscopic
high-temperature limit $T\!\gg\!T_\omega$, since they are of higher
order in $\tanh[\hbar\omega_{kn}/(2k_\mathrm{B}T)]\!\ll\!1$.

\paragraph{Summary.} The demonstrated temperature-indepen\-dence
encountered for molecules with long-wavelength transitions shows that
CP forces on such systems cannot be altered by adjusting the ambient
temperature. Instead, the original zero-temperature results of
Casimir and Polder apply universally across the whole temperature
range. It is worth emphasising that a `classical' regime of linear
temperature-dependence is never reached. Our results further indicate
that when accounting for the thermal excitation of the media, the
temperature-dependence of Casimir forces involving dielectrics may be
weaker than previously thought.

\acknowledgments
We have benefited from discussions with I.~Brevik, A.~Lambrecht and
S.~Reynaud. This work was supported by the UK Engineering and Physical
Sciences Research Council. Support from the European Science
Foundation (ESF) within the activity `New Trends and Applications of
the Casimir Effect' is gratefully acknowledged.


\end{document}